\documentclass[fleqn,usenatbib]{mnras}

\usepackage{newtxtext,newtxmath}

\usepackage[T1]{fontenc}

\DeclareRobustCommand{\VAN}[3]{#2}
\let\VANthebibliography\thebibliography
\def\thebibliography{\DeclareRobustCommand{\VAN}[3]{##3}\VANthebibliography}

\usepackage{graphicx}	
\usepackage{amsmath}	
\usepackage{bbding}
\usepackage{orcidlink}
\usepackage{tikz}
\usepackage{hyperref}

\hypersetup{colorlinks=true}

\title{Kinematical coherence between satellite galaxies and host stellar discs for MaNGA \& SAMI galaxies}
\author[Wang et al.]{
Sen Wang$^{1}$\thanks{E-mail: wangsen19@mails.tsinghua.edu.cn}
\hspace{-1.5mm}$^{~\orcidlink{0009-0007-6358-3564}}$,
Dandan Xu$^{1}$, 
Shengdong Lu$^{2}$
\hspace{-1.5mm}$^{~\orcidlink{0000-0002-6726-9499}}$,
and Cheng Li$^{1}$ \\
$^{1}$Department of Astronomy, Tsinghua University, Beijing, 100084, China\\
$^{2}$Institute for Computational Cosmology, Department of Physics, Durham University, South Road, Durham, DH1 3LE, UK\\
}

\date{Accepted XXX. Received YYY; in original form ZZZ}

\pubyear{2023}

\begin{document}
\label{firstpage}
\pagerange{\pageref{firstpage}--\pageref{lastpage}}
\maketitle  
\begin{abstract}
The effect of angular momentum on galaxy formation and evolution has been studied for several decades. Our recent two papers using IllustrisTNG-100 simulation have revealed the acquisition path of the angular momentum from large-scale environment (satellites within hundreds of kpc) through the circum-galactic medium (CGM) to the stellar discs, putting forward the co-rotation scenario across the three distance scales.
In real observations, although the rotation signature for the CGM and environmental three-dimensional (3d) angular momentum are difficult to obtain, line-of-sight kinematics of group member galaxies and stellar disc kinematics of central galaxies are available utilizing existing group catalogue data and integral field unit (IFU) data. In this paper, we use (1) the group catalogue of SDSS DR7 and MaNGA IFU stellar kinematic maps and (2) the group catalogue 
of GAMA DR4 data and SAMI IFU stellar kinematic maps, to test if the prediction above can be seen in real data. We found the co-rotation pattern between stellar discs and satellites can be concluded with 99.7 percent confidence level ($\sim 3\sigma$) when combining the two datasets. And the random tests show that the signal can be scarcely drawn from random distribution. 
\end{abstract}



\section{Introduction}
\label{sec:introduction}
As one of the most fundamental physical properties, angular momentum plays a key role for us to fully understand the formation and evolution of galaxies. A widely accepted theory explaining the origin of halo angular momentum is the tidal toque theory (TTT) \citep{Peebles_1969,Doroshkevich_1970,White_1984}. The theory suggests that a dark matter halo gains angular momentum at the linear growth stage due to the misalignment between its inertia tensor and the tidal field tensor of the large-scale structure (LSS) around it, resulting in an alignment between the halo's angular momentum and its large-scale environment. When the evolution enters the non-linear regime, the LSS modulation becomes less efficient, gas cools and contracts within the dark matter halo. During this process, it spins up with angular momentum preserved, and becomes denser. Finally star formation is triggered in the gaseous disc and a stellar disc eventually forms. Traditional models assume baryons acquire angular momentum from their dark matter haloes and define a specific angular momentum retention fraction, denoted by $f_{j}$. Usually $f_{j}$ is close to 1 for disc galaxies (e.g. \citealt{Mo_Mao_White_1998,Firmani_Avila-Reese_2000,Firmani_Avila-Reese_2009}), which means baryons possess nearly the same amount of specific angular momentum as dark matter. 
In this sense, as a result, correlations in angular momenta must exist among the large-scale environment and the dark matter halo, the cold circumgalactic medium (CGM), and all the way down to the galaxy's stellar discs.
During the non-linear structure growth, other processes such as gas shock-heating \citep{White_and_Rees_1978,White_and_Frenk_1991} and galaxy mergers, also start to redistribute the angular momenta of the entire system, which tends to erase those correlations.

These correlations have been widely studied using cosmological simulations (e.g., \citealt{Aragon-Calvo_et_al_2007, Hahn_et_al_2007, Libeskind_et_al_2013_a,  Wang_et_al_2018, Lopez_et_al_2021}). 
In particular,  \citet{Moon_et_al_2021} studied the so-called spin-orbit alignment (SOA) of galaxy pairs in IllustrisTNG-100 simulation (\citealt{Marinacci_et_al_2018,Naiman_et_al_2018,Springel_et_al_2018,Nelson_et_al_2018,Nelson_et_al_2019b,Pillepich_et_al_2018b,Pillepich_et_al_2019}). They found significant alignment between the pair orbital angular momentum and the central star, gas, and DM spins, which only exists in central-satellite pairs, and the closer the stronger.
Using cosmological N-body simulations, \citet{An_et_al_2021} concluded that such SOA is possibly resulted from the local cosmic flow along the filament with a further growth due to neighbouring interactions. 
A recent paper series \citep{Wang_et_al_2022,Lu_et_al_2022} using the Illustris TNG-100 simulation revealed the modulation of a galaxy's ambient angular momentum environment (i.e. the satellite galaxies) on the central star formation activeness, via the cold CGM gas. They demonstrated a significant signal of coherent rotation among the stellar disc, the cold CGM, and satellite galaxies (see Figure 9 of 
\citealt{Lu_et_al_2022}). Similar coherent signals between spins of the DM halo and those from their neighbours were also confirmed by \citet{Kim_et_al_2022} using N-body simulations. Observationally, great efforts have also been made searching for correlated signals between galaxies/halos and their LSS environment (e.g. \citealt{Lee_Erdogdu_2007, Kraljic_et_al_2021, Tudorache_et_al_2022, Barsanti_et_al_2022}). In particular, kinematic coherence has been reported to exist with large significance using a sample of CALIFA galaxies in their large-scale environment out to several Mpc (e.g., \citealt{Lee_et_al_2019a,Lee_et_al_2019b}). 

In this paper, we follow the predictions of \citet{Wang_et_al_2022, Lu_et_al_2022} and investigate the correlations between a galaxy's stellar spin and the kinematics of the orbit motion of satellite galaxies in its neighborhood. In particular we search the signal 
using group catalogues of SDSS and GAMA surveys, and the corresponding IFU observations therein, i.e., the MaNGA and SAMI, respectively. The paper will be organized as follows. In Section~\ref{samples}, we present the galaxy samples and the selection criteria. In Section~\ref{method}, we describe the methods. In Section~\ref{result}, the final results are shown. Finally, summaries and conclusions are presented in Section~\ref{Conclusions_Discussion}. The involved calculations in following sections adopt different cosmological parameters, with the tot matter density of $\Omega_{\rm m} = 0.238$, the cosmological constant of $\Omega_{\Lambda} = 0.762$, and the Hubble constant $h \equiv H_0/(100\,{\rm km s}^{-1} {\rm Mpc^{-1}}) = 0.73$ for SDSS and MaNGA galaxies, and $\Omega_{\rm m} = 0.3$, $\Omega_{\Lambda} = 0.7$, $h = 0.7$ for the GAMA and SAMI galaxies.

\section{Galaxy Samples}
\label{samples}
\subsection{The SDSS and MaNGA}
\label{SDSS}
The Sloan Digital Sky Survey (SDSS, \citealt{York_et_al_2000}) is a large survey aiming at mapping a large field of sky by capturing images and taking spectra of distant objects. 
Until now, the SDSS has produced spectroscopic data for millions of galaxies and photometric data for even a lot more. A number of catalogues have been built up based on these data productions for different scientific goals. In this paper, we adopt the group catalogue constructed by \citet{Yang_et_al_2007} (hereafter YangDR7) using their own halo-based group finder. Applying this group finder to the New York University Value-Added Galaxy Catalog Data(NYU-VAGC, \citealt{Blanton_et_al_2005}) based on SDSS DR7 \citep{Abazajian_et_al_2009}, they obtain 473,872 galaxies and groups, among which 67,925 contain at least two Friends-of-Friends (FoF) members and 44,441 are galaxy pairs (Here we just simply refer the groups with two members as the pairs). 
Throughout this paper, we refer to the brightest galaxies as the central galaxies of the groups, and all other member galaxies as satellites. 

As one of the fourth generation of SDSS (SDSS-IV, \citealt{Blanton_et_al_2017}), Mapping Nearby Galaxies at Apache Point Observatory (MaNGA, \citealt{Bundy_et_al_2015}) has accomplished its task on mapping the kinematic structure and detailed composition of more than 10,000 nearby galaxies with stellar mass $\log M_{\ast}/\mathrm{M_{\odot}} > 9.0$ and redshifts range of $0.001 < z < 0.15$. MaNGA has obtained spatially resolved spectra within the central 1.5-2.5 effective radii of galaxies with 1-2 kpc physical sampling \citep{Drory_et_al_2015}, thanks to its 17 integral field units (IFUs). It provides a wavelength coverage of 360$-$1000 nm and a spectral resolution of $R \sim 2000$ \citep{Bundy_et_al_2015}. 
These spectra provide us with two dimensional maps of stellar kinematics, star formation rate (SFR), stellar metallicity, and so on. The observed $\sim$ 10000 galaxies spread across $\sim$ 2700 $\mathrm{deg^2}$ of the sky. No special selection criteria on apparent size or inclination are made, so these galaxies are representative samples in the local universe. In this paper, we used the stellar kinematic maps from the newest Data Analysis Pipeline (DAP, \citealt{Westfall_et_al_2019,Belfiore_et_al_2019}) output in Data Release 17 (DR17, \citealt{Abdurrouf_et_al_2022}) and the corresponding summary table called the DAPall catalogue. The $\tt DAPTYPE$ (see MaNGA analysis pipeline website) of the kinematic maps we used was labelled as $\tt VOR10-MILESHC-MASTARSSP$, meaning that the map spaxels are binned to $S/N \sim 10$ using the Voronoi binning algorithm \citep{Cappellari_et_al_2003}, that the stellar kinematics are determined using the $\tt MILESHC$ template library \citep{Westfall_et_al_2019}, and that the stellar continuum are fitted by the $\tt MASTARSSP$ templates \citep{Abdurrouf_et_al_2022}. After matching the MaNGA galaxies with YangDR7, we obtain 2429 central galaxies which have at least one satellite member. Among them, 2428 centrals are labeled with morphological types obtained from the MaNGA Deep Learning DR17 Morphology catalogue \citep{Dominguez_et_al_2022} and the distribution of galaxy types is shown in upper panel of Fig.\,\ref{fig:Galaxy_Type}. To avoid edge effect, which accounts for the fact that a group may partly fall out of the survey edges, but not to reduce our sample size too much, we give a cut to let the edge factor larger than 0.9 (edge factor of 1 means no edge effect), leaving 688 LTGs and 133 S0s.

\subsection{The GAMA and SAMI}
\label{GAMA}
The heart of the Galaxy And Mass Assembly (GAMA, \citealt{Driver_et_al_2011}) project is a spectroscopic survey aiming at measuring $\sim$ 300,000 galaxies 
in detail using the AAOmega multi-object spectrograph on the Anglo-Australian Telescope (AAT). The project is divided into three distinct phases, GAMA I \citep{Baldry_et_al_2010}, GAMA II \citep{Liske_et_al_2015,Baldry_et_al_2018}, and GAMA III (GKV, \citealt{Bellstedt_et_al_2020,Driver_et_al_2022}). Until now, the GAMA team has released the fourth and final GAMA data release (DR4, \citealt{Driver_et_al_2022}) which will be updated when new versions of data are available. In this paper, we used the catalogues of GAMA II in GAMA DR4 for our analysis. The GAMA II catalogues cover three different survey regions: equatorial,  G02, and G23 (see GAMA webpage\footnote{\url{http://www.gama-survey.org/dr4/}} for details), containing $\sim$ 460,000 science targets. The GAMA data is arranged as the so-called Data Management Units (DMUs), each of which contains a particular type of catalogue data or tables. The {\sc GroupFinding} DMU \citep{Robotham_et_al_2011} provides group catalogues for GAMA galaxies in equatorial and G02 regions, giving 26,194 groups with at least two FoF members out of 83,093 galaxies. 

\begin{figure}
\centering
\includegraphics[width=1\columnwidth]{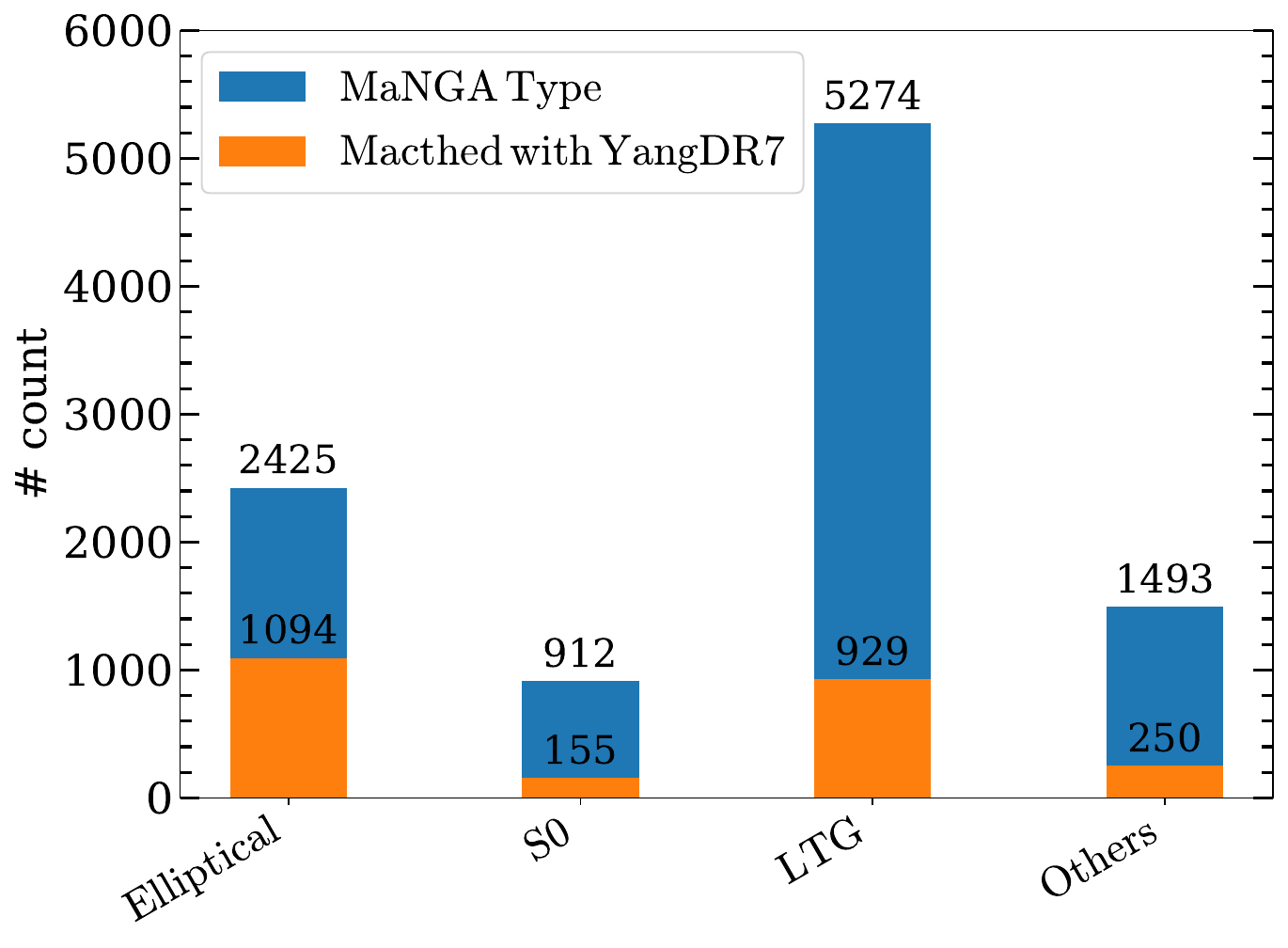}
\includegraphics[width=1\columnwidth]{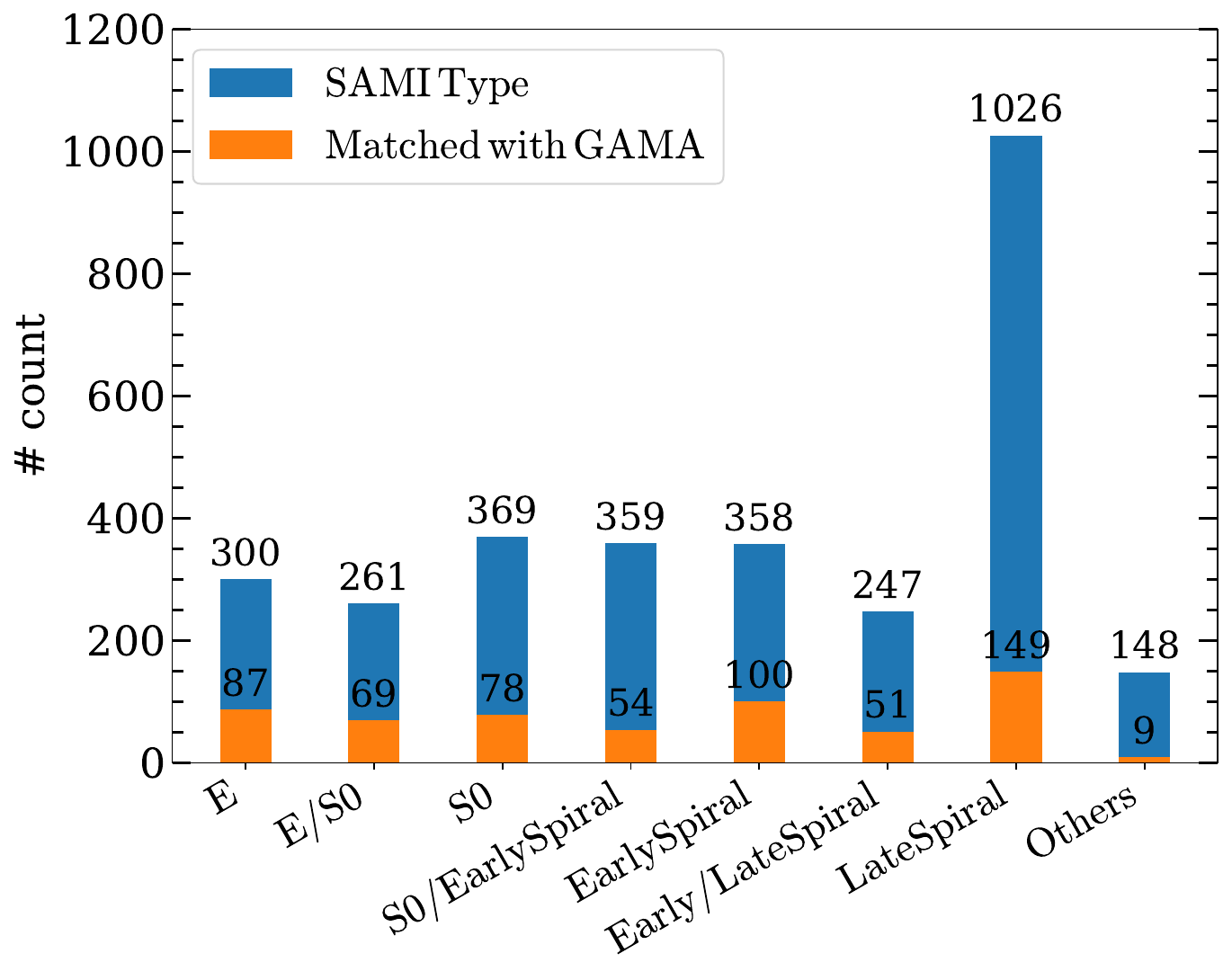}
\caption{The statistics of galaxy types of MaNGA (upper panel) and SAMI (lower panel) for all galaxies (blue) and matched centrals (orange). The x-axis shows the different galaxy morphology types defined in each survey and the numbers on the histograms are the corresponding counts.}
\label{fig:Galaxy_Type}
\end{figure}

The Sydney-Australian-Astronomical-Observatory Multi-object Integral-Field Spectrograph (SAMI, \citealt{Croom_et_al_2012}) is a new instrument on the 4-meter Anglo-Australian Telescope equipped with integral-field spectroscopy (IFS) instrument, which allows us to spatially measure galaxies in great detail. Recently, the newest SAMI data release three (DR3, \citealt{Croom_et_al_2021}) containing 3068 galaxies and their value-added data products (such as visual morphology) have been available. These galaxies have stellar masses ranging from $10^{8}\mathrm{M_{\odot}}$ to $10^{12}\mathrm{M_{\odot}}$ with redshift range of $0.004<z<0.115$. The SAMI instrument has 13 so-called hexabundles, each of which contains 61 optical fibres with each subtending 1.6 arcsec, giving a total diameter of 15 arcsec for one bundle \citep{Bryant_et_al_2014}. The angular resolution of SAMI is expected to be $\sim$ 2.1 arcsec considering seeing.
For spectral sampling, SAMI has a resolution of $R\sim 1700$ in the blue arm with a wavelength coverage of 3700–5700 {\AA} and a resolution of $R\sim 4500$ in the red arm covering a wavelength range of 6300–7400 {\AA} \citep{Bryant_et_al_2015}. The stellar kinematics in SAMI are derived using penalized Pixel-Fitting code (pPXF, \citealt{Cappellari_Emsellem_2004,Cappellari_2017}). Matching SAMI galaxies with the GAMA group catalogue 
we obtained 597 groups, where 432 centrals are spirals/S0s. The morphologies of all SAMI galaxies are visually classified (see \citealt{Cortese_et_al_2016} for details) by the team through SDSS or VST/ATLAS RGB composite images and are available in {\sc VisualMorphologyDR3} catalogue. The counts of different types are shown in the lower panel of Fig.\,\ref{fig:Galaxy_Type}. Among all of the centrals, 562 galaxies have well-defined kinetic position angles (PAs). Criterion that {\tt GroupEdge} value of groups shall be larger than 0.9 is also applied, leaving 400 spirals/S0s.

\begin{figure}
\centering
\includegraphics[width=1\columnwidth]{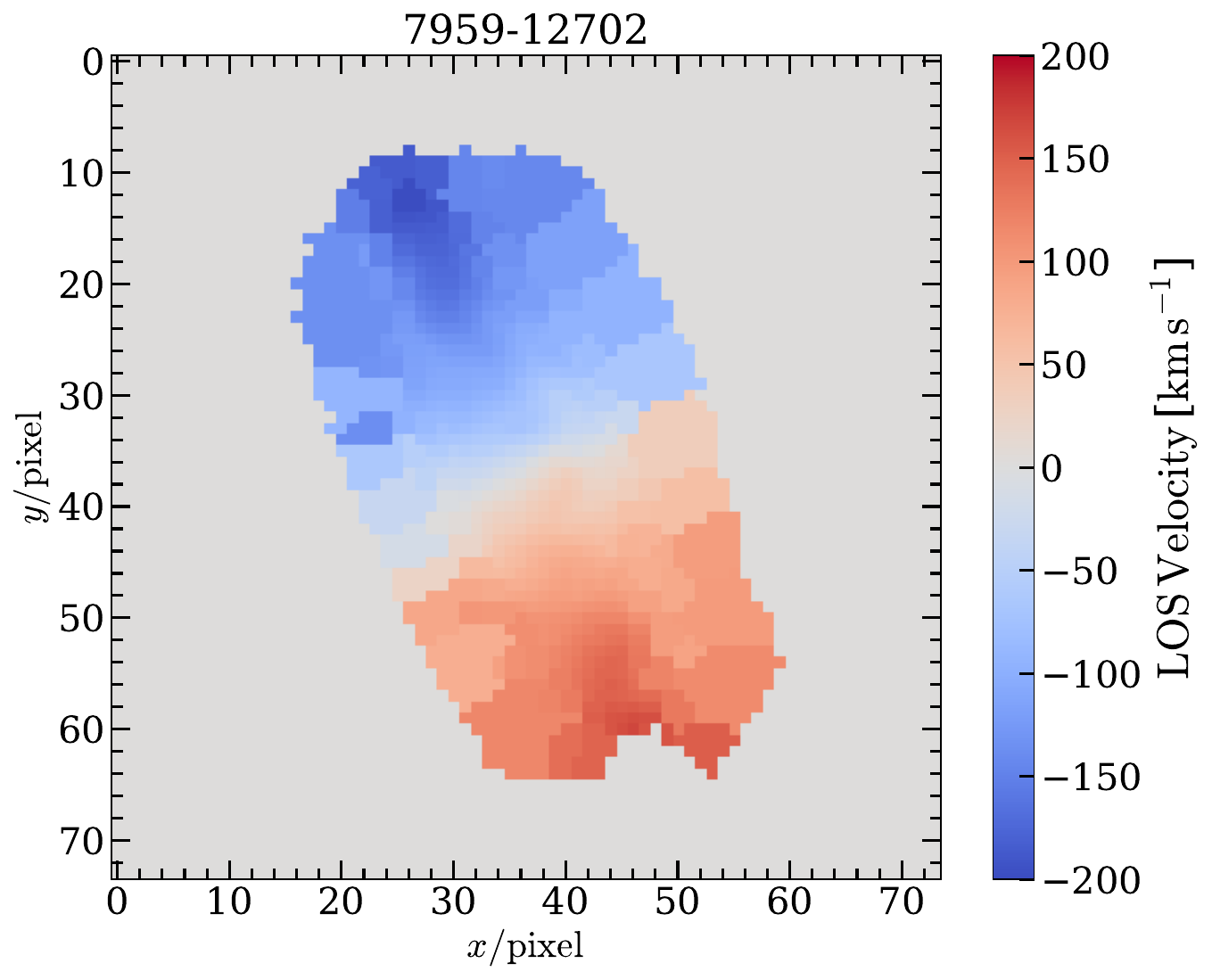}\\
\includegraphics[width=1\columnwidth]{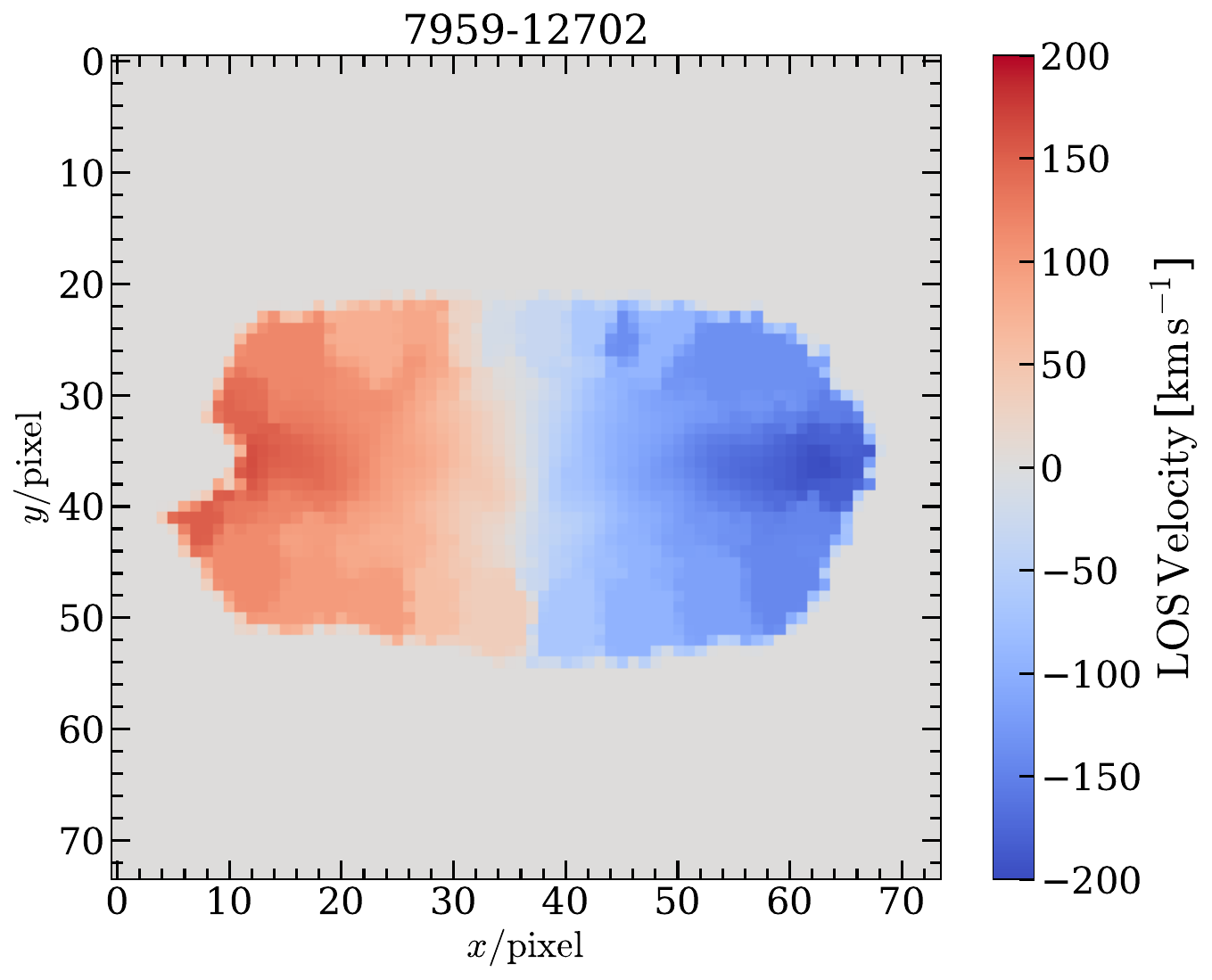}
\caption{An example (from MaNGA, Plate-IFU: 7959-12702) on the rotation of the stellar discs. We rotate the original kinematic map (top panel) such that the kinetic major axis of the disc is then along the x-axis (bottom panel). We always make sure that the red-shifted (blue-shifted) part lies on the left (right) side of the x-axis. The color shows the line-of-sight velocities of pixels.}
\label{fig:PA_example}
\end{figure}

\begin{figure*}
\centering
\includegraphics[width=1\columnwidth]{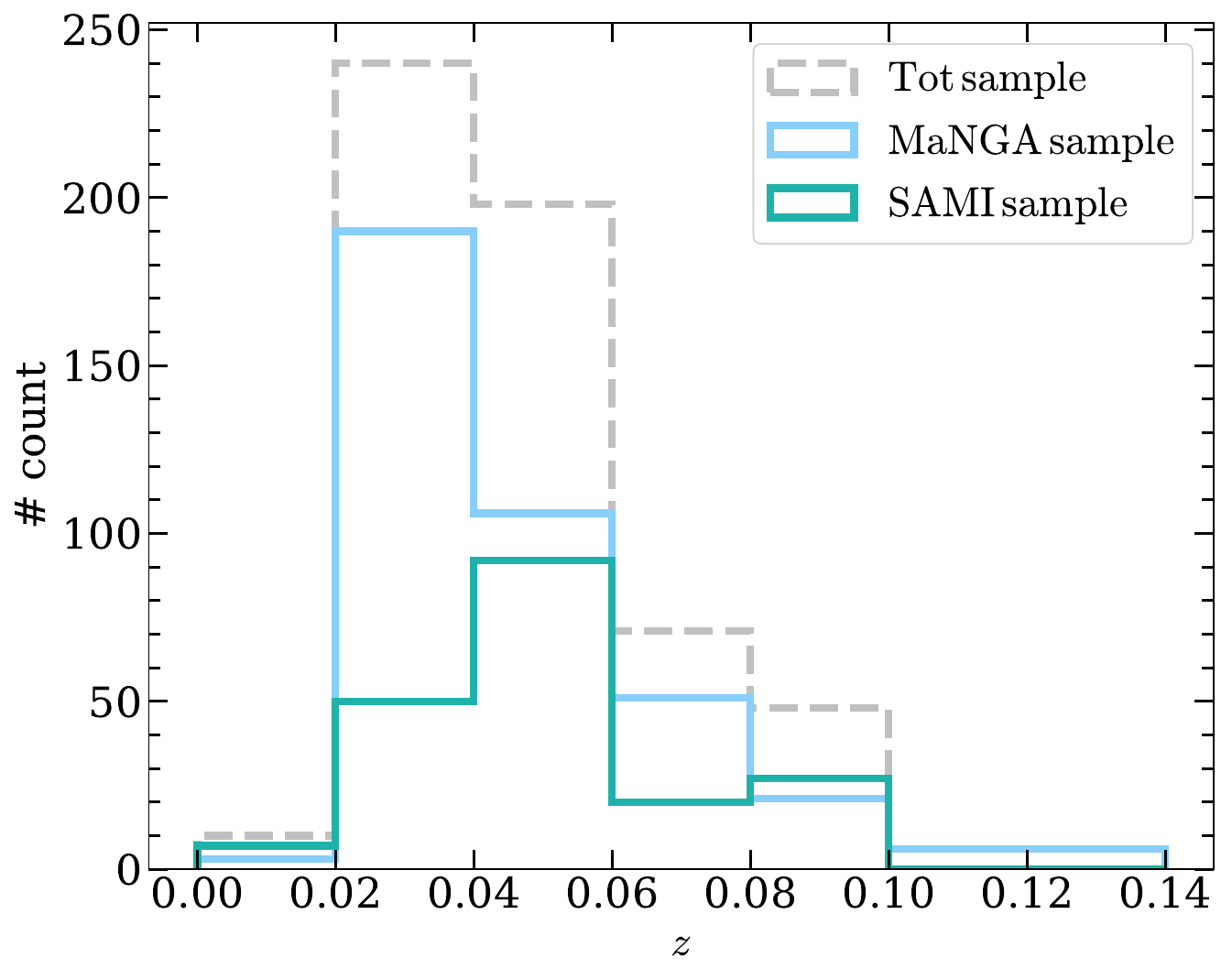}
\includegraphics[width=1\columnwidth]{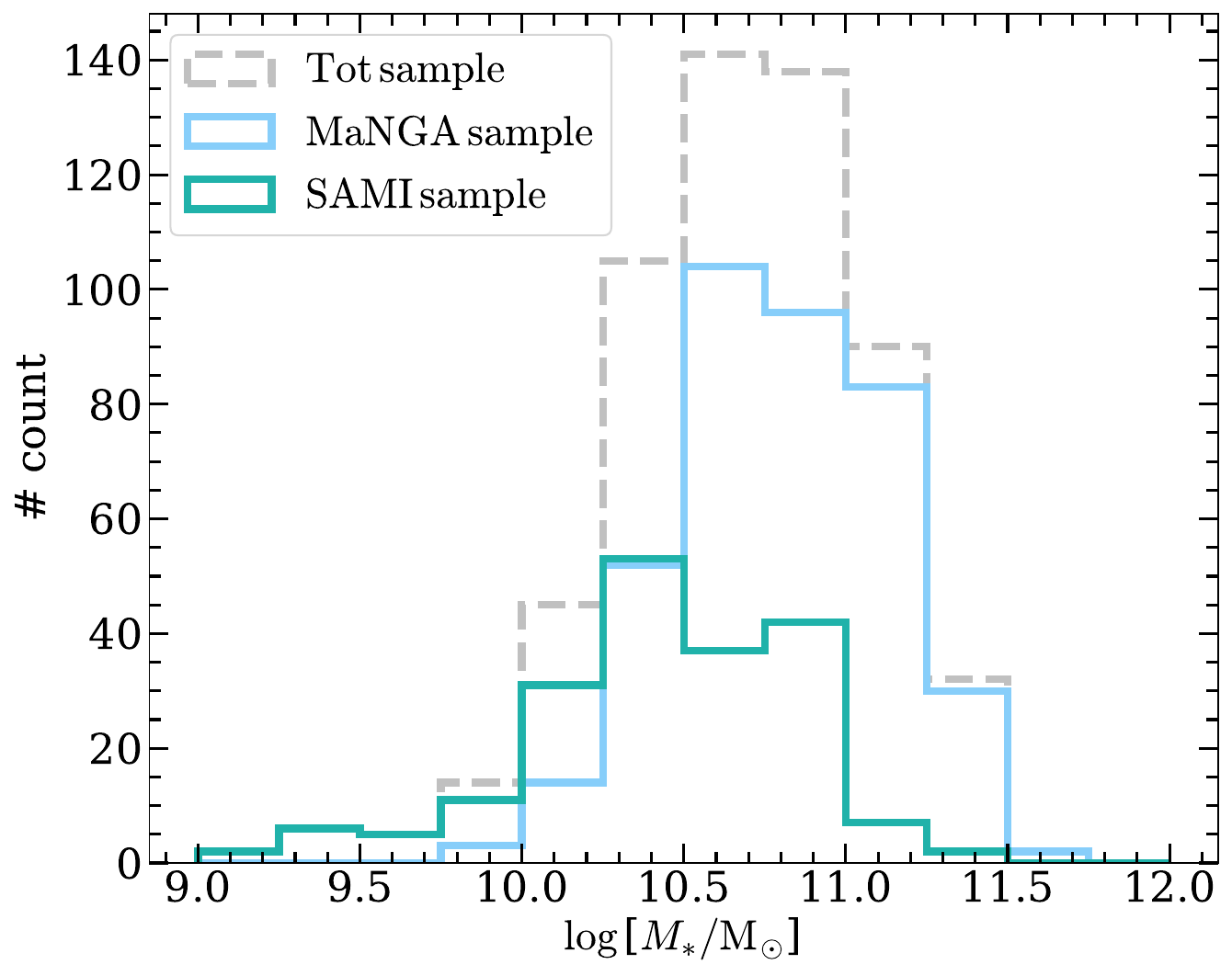}
\includegraphics[width=1\columnwidth]{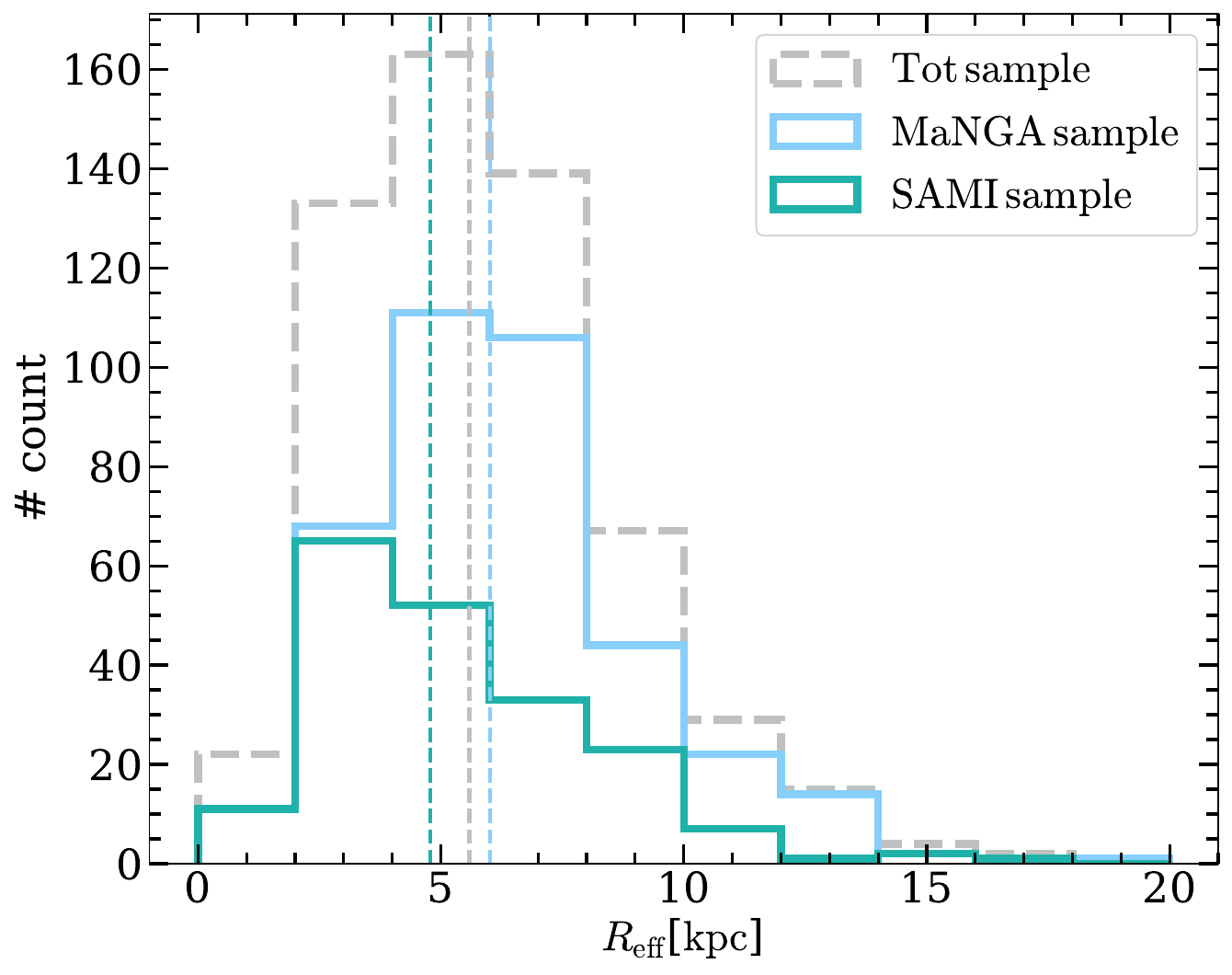}
\includegraphics[width=1\columnwidth]{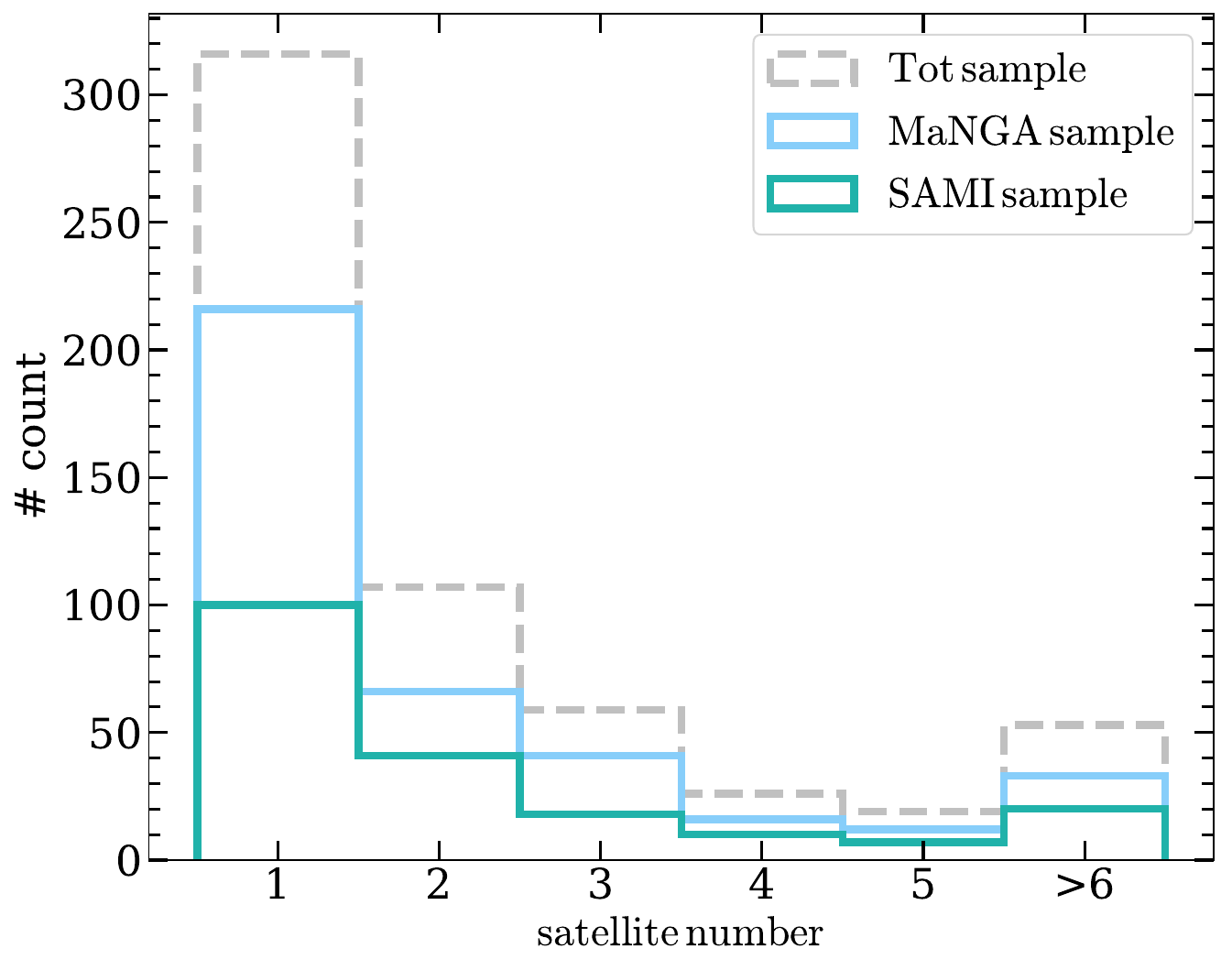}
\caption{The distribution of central redshift, central stellar mass, central effective radius, and the satellite number of the selected samples. Silver, blue, and green lines are for tot samples, MaNGA samples, and SAMI samples respectively. Vertical lines in the lower left panel represent the medians.}
\label{fig:Central statistic}
\end{figure*}

\section{Methodology}
\label{method}

\subsection{Stacking the kinetic fields of host stellar discs and satellites }
\label{Stack}
In Figure 9 of \citet{Lu_et_al_2022}, the authors demonstrated the co-rotation signals by stacking the line-of-sight velocity maps of the ambient cold CGM gas and satellites of TNG-100 galaxies according to their corresponding stellar discs.
The steps therein are summarized as follows: first all galaxy discs were rotated to their edge-on configurations using the corresponding rotational matrices; secondly a random disc inclination between $-60^{\circ}$ and $60^{\circ}$ was applied to each simulated galaxy around the major axis considering that observed galaxies are rarely perfectly edge-on. The red-shifted sides of galaxy discs were all put on the same side in the sky plane during stacking. Finally, the same operations were done for all satellite galaxies and the CGM gas cells. After stacking, a coherent pattern in the line-of-sight kinematics appears among the galaxy discs, the associated cold CGM gas, and the satellite galaxies in the environment. 

When dealing with the collected observational data in this work, we took a similar approach but with a few differences due to the limitation of the data. Both MaNGA and SAMI provide catalogues of morphological type of galaxies, based on which we first selected late type galaxies (LTG) and S0 galaxies for MaNGA and all spirals and S0 galaxies for SAMI as our stacking samples because of their well-recognized stellar discs. As we have no idea about the rotation matrices of galaxies in their sight lines, so we selected galaxies that appear nearly edge-on with clear rotation kinematic (see Fig.\,\ref{fig:PA_example}). We note that although MaNGA also gives a parameter in its morphological catalogue describing whether the galaxy is edge-on, the edge-on criterion is too strict to result in any sizable statistical sample (113 LTGs/25 S0s in 688 LTGs/133 S0s). We therefore did NOT adopt the MaNGA edge-on parameter as indication. Instead, we took a visual inspection approach through images and two-dimensional (2D) velocity maps, and finally obtained 326 LTGs and 58 S0s for MaNGA galaxies, and 196 spirals/S0s for SAMI galaxies that have kinematic maps with good sense of rotations. The statistics (i.e. redshift, stellar mass, effective radius, and satellite numbers) of the selected systems are shown in Fig.\,\ref{fig:Central statistic}. The stellar masses of the selected galaxy samples span a range of $\log M_{\ast}/\mathrm{M_{\odot}} \in [9.0, \,11.5]$. The median effective radii of the two galaxy sample are about 5 kpc. Most of the systems have one or two satellites. 

We then took the kinematic PA of each central galaxy to rotate the overall kinematic field. The adopted PAs for MaNGA galaxies were derived from \citet{Zhu_et_al_2023}, which adopted the \textsc{PaFit}\footnote{Version 2.0.7, from \url{https://pypi.org/project/pafit/}} package
and the PAs for SAMI galaxies were calculated using the $\tt FIT\_ KINEMATIC\_ PA$ code. Both of them are derived using the algorithm described in Appendix C of \citet{Krajnovic_et_al_2006}.

The kinematic PA of each central galaxy was then used to rotate the stellar kinematic map, as well as the associated satellites galaxies such that the projected major axis is along the x-axis as is presented in Fig.\,\ref{fig:PA_example}. A further rotation was given to make the red-shifted (blue-shifted) side of the stellar kinematic map appear on the left-hand (right-hand) side during stacking. 

As the galaxy sample spans a stellar mass range over two orders of magnitudes, a further normalization on the projected separation is required before stacking in order to eliminate the mass dependence. To do so, we scaled the physical distances along the projected major axis with the corresponding effective radius $R_{\rm eff}$, and those along the minor axis by $\frac{b}{a} R_{\rm eff} $, where $\frac{b}{a}$ is the projected axis ratio of the central galaxy.

\subsection{Distinguishing the red- and blue-shifted satellites}
\begin{figure}
\centering
\includegraphics[width=1\columnwidth]{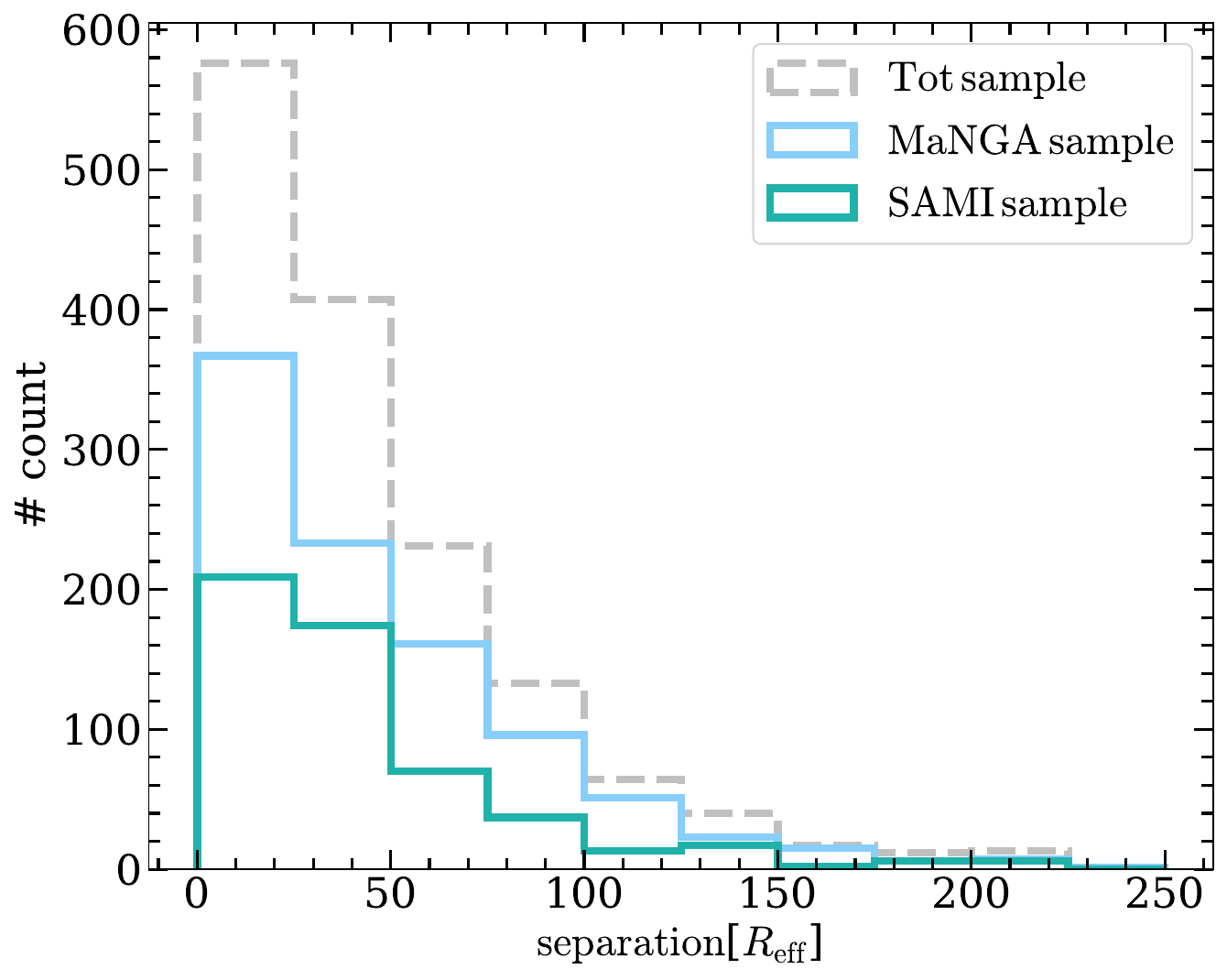}
\includegraphics[width=1\columnwidth]{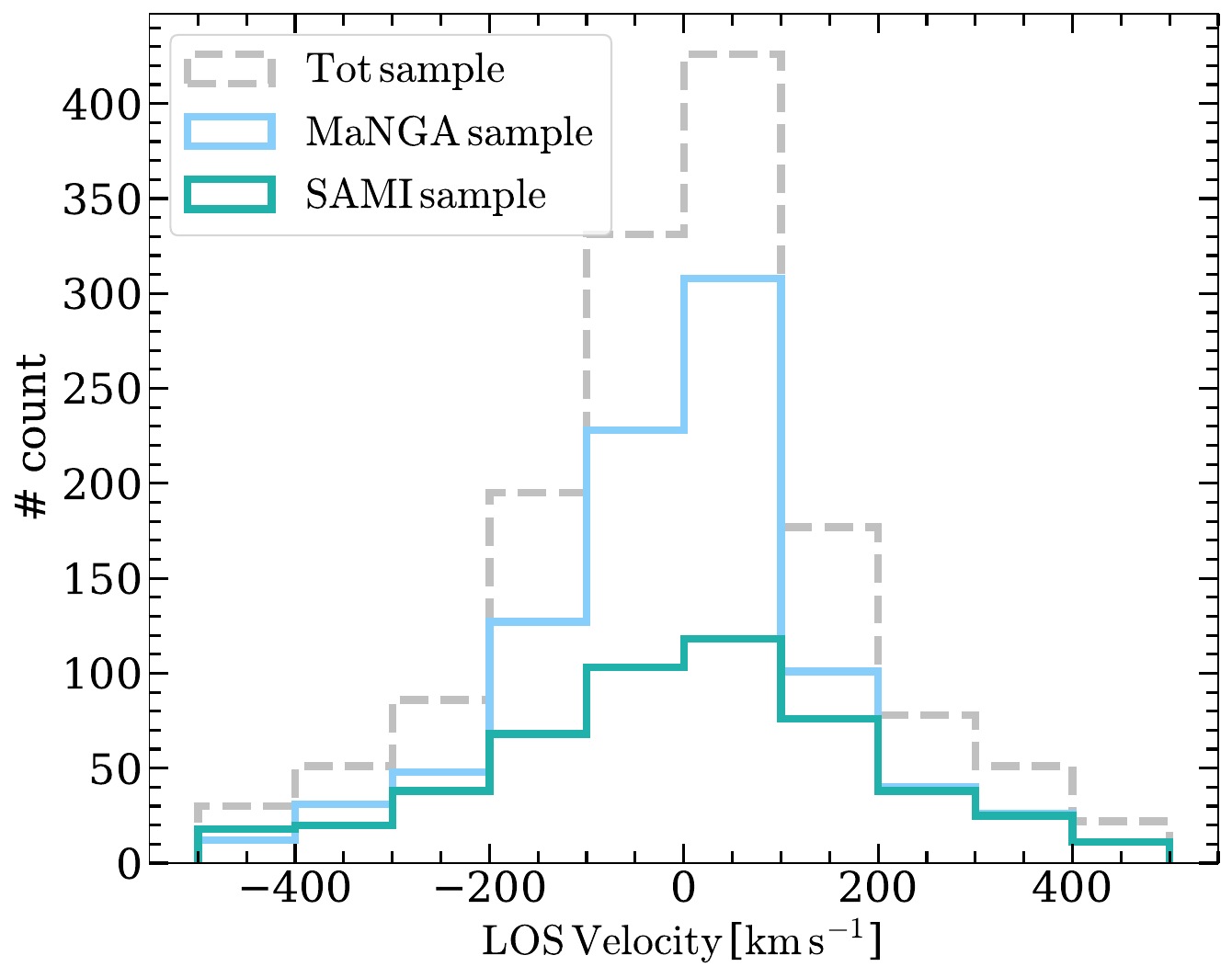}
\caption{Histograms of the satellite projected separations (top) and LOS velocities (bottom) relative to the central galaxies.}
\label{fig:Satellite statistic}
\end{figure}

Considering the spectral resolution and the fact that satellites with too high relative velocities are not likely to be bound to the systems, we further retained satellites with relative line-of-sight (LOS) velocities (see below for details) ranging from $30$ km/s to $500$ km/s. The line-of sight velocities of satellites relative to their centrals are calculated based on the assumptions that the centrals present no peculiar motions and the peculiar motions of satellites are non-relativistic. The relative line-of sight velocity is defined as:
\begin{equation}
\label{LOS}
\Delta v_{\rm los} = \frac{z_{s}-z_{c}}{1+z_{c}}\, c
\end{equation}
where $z_{c}$ and $z_{s}$ are the redshifts of the central and satellite galaxies, respectively; $c$ is the speed of light. The statistics of the satellite projected separations and LOS velocities are presented in Fig.\,\ref{fig:Satellite statistic}

If satellite galaxies indeed exhibit coherent kinematics as the central stellar discs, one would expect that the red-shifted (blue-shifted) satellites shall be preferentially distributed on the red-shifted (blue-shifted) side of the stellar discs of the host galaxies and therefore the stacked (normalized) separation distribution along the major axis of the red-shifted satellite galaxies shall be significantly different from that of the blue-shifted satellite population. To quantitatively describe this difference, we applied Kolmogorov-Smirnov (K-S) test to the two distributions using the {\sc python} package named NumPy\footnote{\url{https://numpy.org}}, which provides us a $p$-value, indicating the probability that the two distributions come from the same underlying distribution. The lower $p$ is, the larger the difference is between the two distributions.

We also ran random tests to our samples. We maintained the radial ordering of the stacked distribution but assigned a random angle relative to the major axis to each satellite, resulting in one particular realization of random angular distribution. Then we did K-S test, giving us a $p$-value. After repeating this sampling and testing process for enough times (e.g. 100,000 times), we obtained a group of distributions and $p$-values. The 1 $\sigma$ regions of these distributions along major axis and the median with the standard deviation of the $p$-values are taken as the result of the random tests.

The other way to test the coherent kinematics is that if the coherent signal exists, the satellites that satisfy $x<0$ (i.e., on the red-shifted side of the stellar discs) would preferentially exhibit red-shifted LOS velocities; those with $x>0$ would prefer blue-shifted LOS velocities. We quantified the difference between the two LOS velocity distributions also using the K-S test. Again, a smaller value of $p$ indicates a lower chance of the two distributions are intrinsically the same.

\section{Results}
\label{result}
\label{Co-rotation}
\begin{figure}
\centering
\includegraphics[width=1\columnwidth]{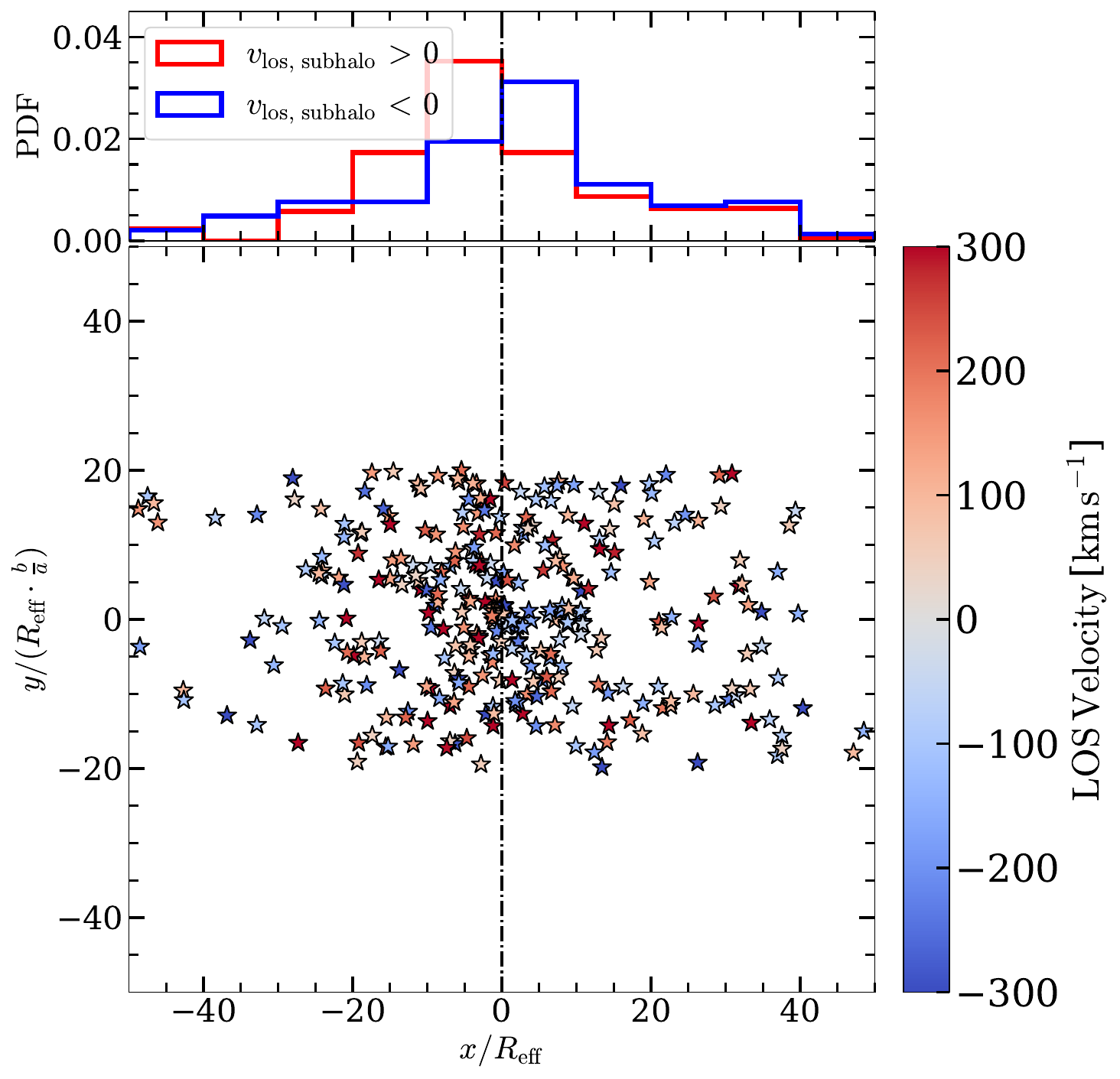}
\caption{The stacked distribution of satellites color-coded by the line-of-sight velocities relative to their corresponding central galaxies. All central galaxies are rotated according to their kinetic PAs such that the kinetic major axes are all along the x-axis; and the red- and blue-shifted side are put to the left- and right-hand side of the figure, respectively. The x- and y-axis are the normalized projected physical scale calculated at the central's redshift with the corresponding cosmology (see Sec.\,\ref{sec:introduction}). The star symbols represent the collections of the satellites, which are rotated accordingly and that meet the criteria of $|y|\,<\,20\,R_{\rm eff}\cdot \frac{b}{a}$. The histograms on the top of the panel show the distribution of these satellites along the major axis, with the red and blue lines for the positive and negative line-of-sight velocities respectively. Black dash-dotted line marks the $x=0$.}
\label{fig:MaNGA_SAMI}
\end{figure}

According to the structure formation and evolution theory, there should be expected correlations between large and small scales as have been seen in various simulations (e.g. \citealt{Aragon-Calvo_et_al_2007,Hahn_et_al_2007,Libeskind_et_al_2013_a,Libeskind_et_al_2013_b,Wang_et_al_2018,Lopez_et_al_2021}) and observations (e.g. \citealt{Lee_Erdogdu_2007,Lee_et_al_2019a,Lee_et_al_2019b,Kraljic_et_al_2021,Tudorache_et_al_2022,Barsanti_et_al_2022}). Here we study the coherent dynamics between the central discs and their satellites by stacking all samples, as is shown in Fig.\,\ref{fig:MaNGA_SAMI}. From this relative spatial configuration, 
we see that the red-shifted (blue-shifted) satellites are preferentially distributed on the red-shifted (blue-shifted) side of the stellar discs of the host galaxies, and a significant difference appears between the distributions of satellites along the major axis (i.e. red and blue histograms), especially at scales within $\pm 20\,R_{\rm eff}$, about $\pm 100$ kpc. This difference indicates a preferential coherent rotation between satellite galaxies and the stellar discs of host galaxies and the scale is totally comparable with the simulated results from the TNG-100 simulation, as is shown in the Figure 9 of \citet{Lu_et_al_2022}. 

\begin{figure}
\centering
\includegraphics[width=1\columnwidth]{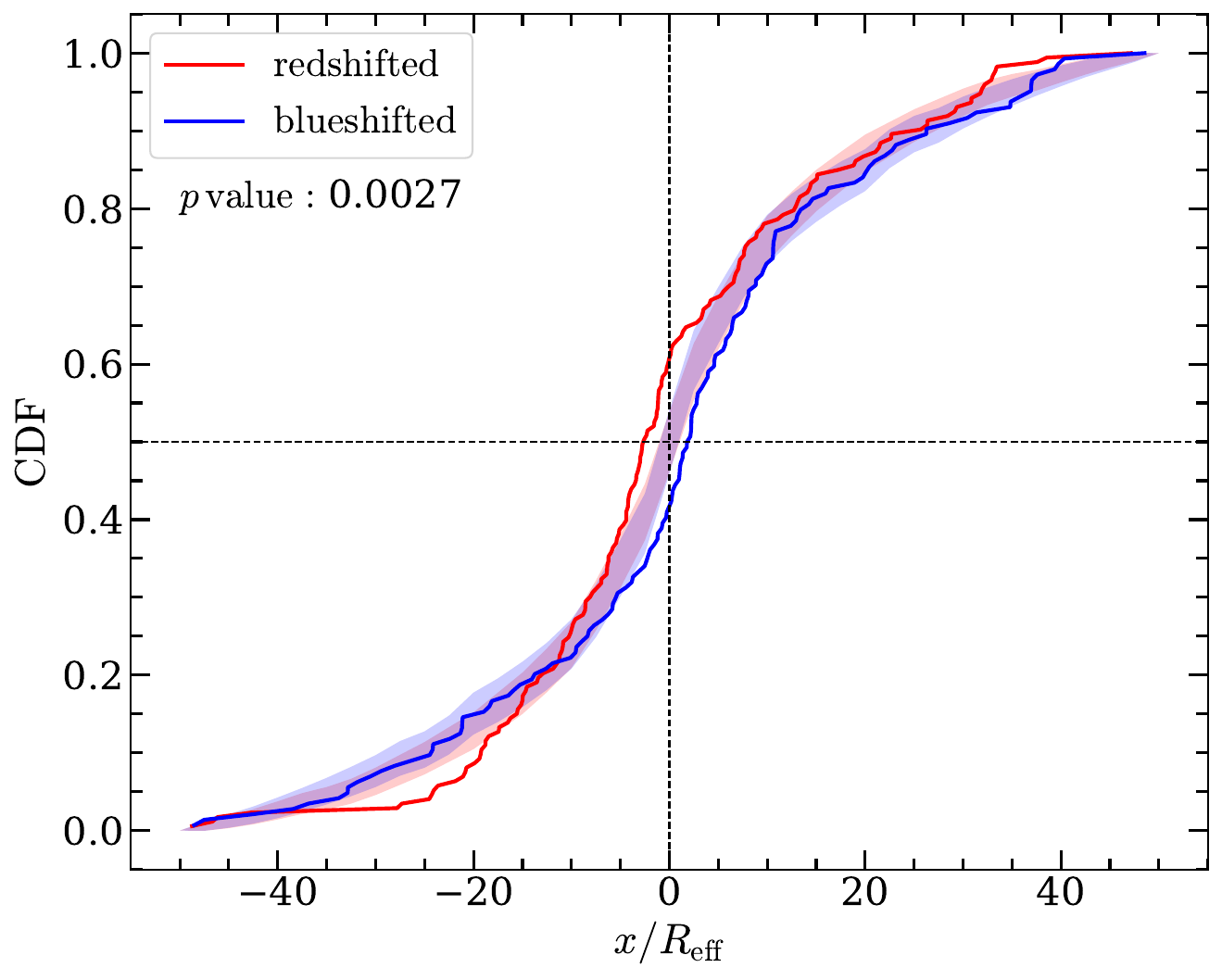}
\caption{The cumulative distribution functions (CDFs) along the major axis of the satellites in Fig.\,\ref{fig:MaNGA_SAMI} for all red-shifted and blue-shifted members (red and blue lines) but restricted to the same region. Solid lines are the distributions for real data while shaded parts are the $1\sigma$ regions of the 100,000 random angular sampling tests.
Dashed vertical and horizontal lines represent the zero point along the major axis and CDF value of 0.5. On the upper left shows the $p$-value of the K-S test between blue-shifted and red-shifted species.}
\label{fig:K-S_test}
\end{figure}

In order to quantify the differences between these two distributions, we applied a Kolmogorov–Smirnov test and the result is displayed in Fig.\,\ref{fig:K-S_test}.
The blue and red solid lines are the cumulative distribution functions (CDFs) of the projected separation along the major axis for the blue-shifted and red-shifted satellite galaxies, respectively. 
The dashed vertical and horizontal lines show the zero point on the x-axis and the 50-percent accumulation points of the CDF. 
The CDF value of blue-shifted satellites (blue solid line) is about 0.43 at $x=0$, indicating an asymmetric distribution toward the blue-shifted part of the stellar discs. The same but more asymmetric signal is seen for the red-shifted satellites, for which the CDF at $x=0$ is $\sim 0.62$. 
Through a K-S test, we found a p-value of $\sim$ 0.0027, indicating that the hypothesis that the two distributions are the same can be ruled out at $\sim 3\sigma$ level, or in other words, that the red-shifted and blue-shifted data are totally different distributions at a confidence level of at least 99.7 percent. The fact that the red-shifted (blue-shifted) satellites preferentially projected on the $x<0$ ($x>0$) side, implies the expected coherent rotation between satellites and the central discs. The $1\sigma$ regions (shaded) of the random tests show nearly no preferential distributions and the median of the $p$-values ($\sim 0.56_{-0.36}^{+0.31}$) indicates that this kinematical coherence can hardly be drawn from the random distributions.

\begin{figure}
\centering
\includegraphics[width=1\columnwidth]{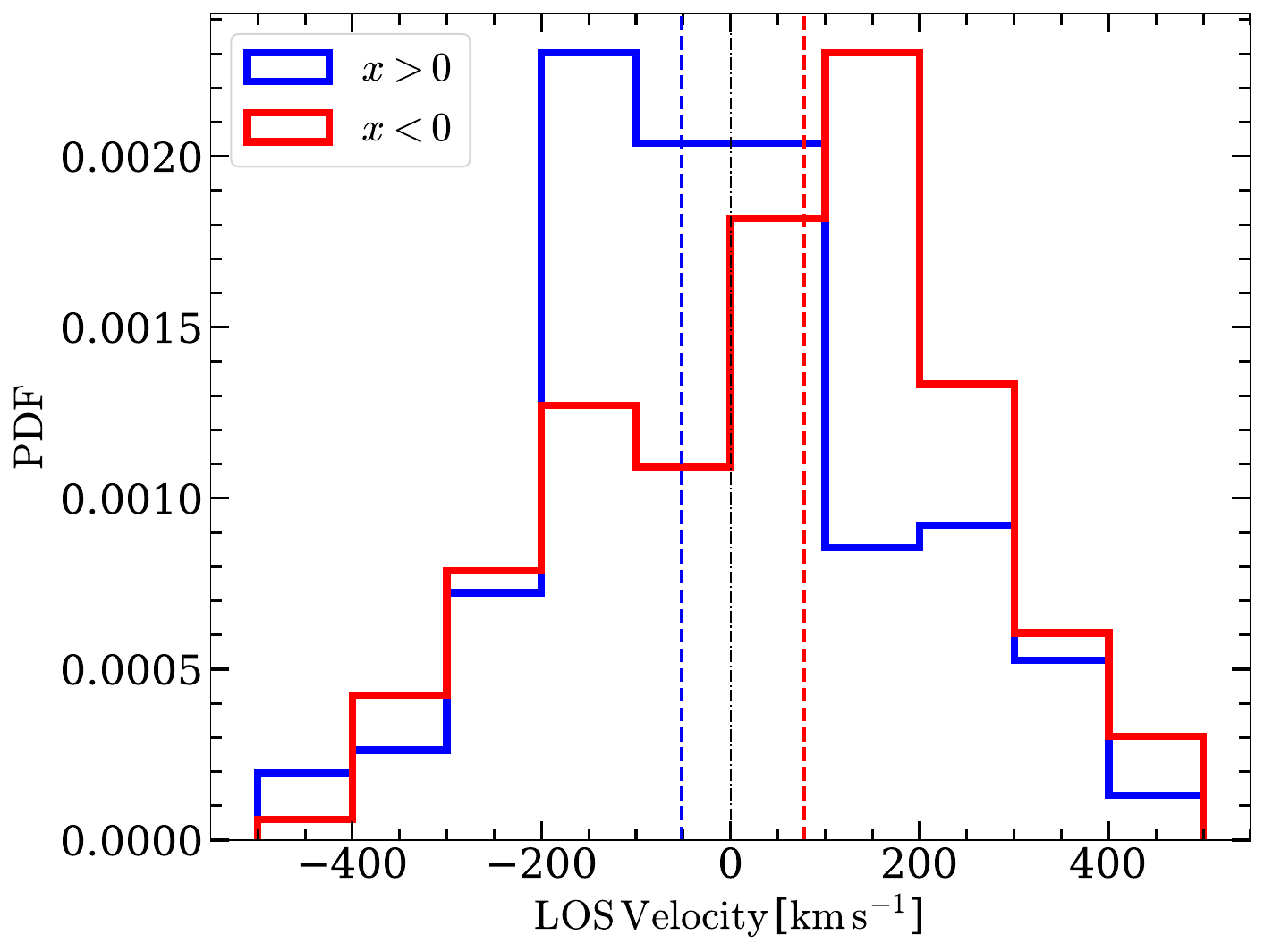}
\caption{LOS velocity distributions of the satellites shown in Fig.\,\ref{fig:MaNGA_SAMI}. Blue and red solid lines are the histograms for satellites with $x>0$ and $x<0$ respectively while dashed lines are the corresponding medians of the distributions. The black dash-dotted line just indicates the LOS velocity of zero. }
\label{fig:LOS_distribution}
\end{figure}

\begin{figure}
\centering
\includegraphics[width=1\columnwidth]{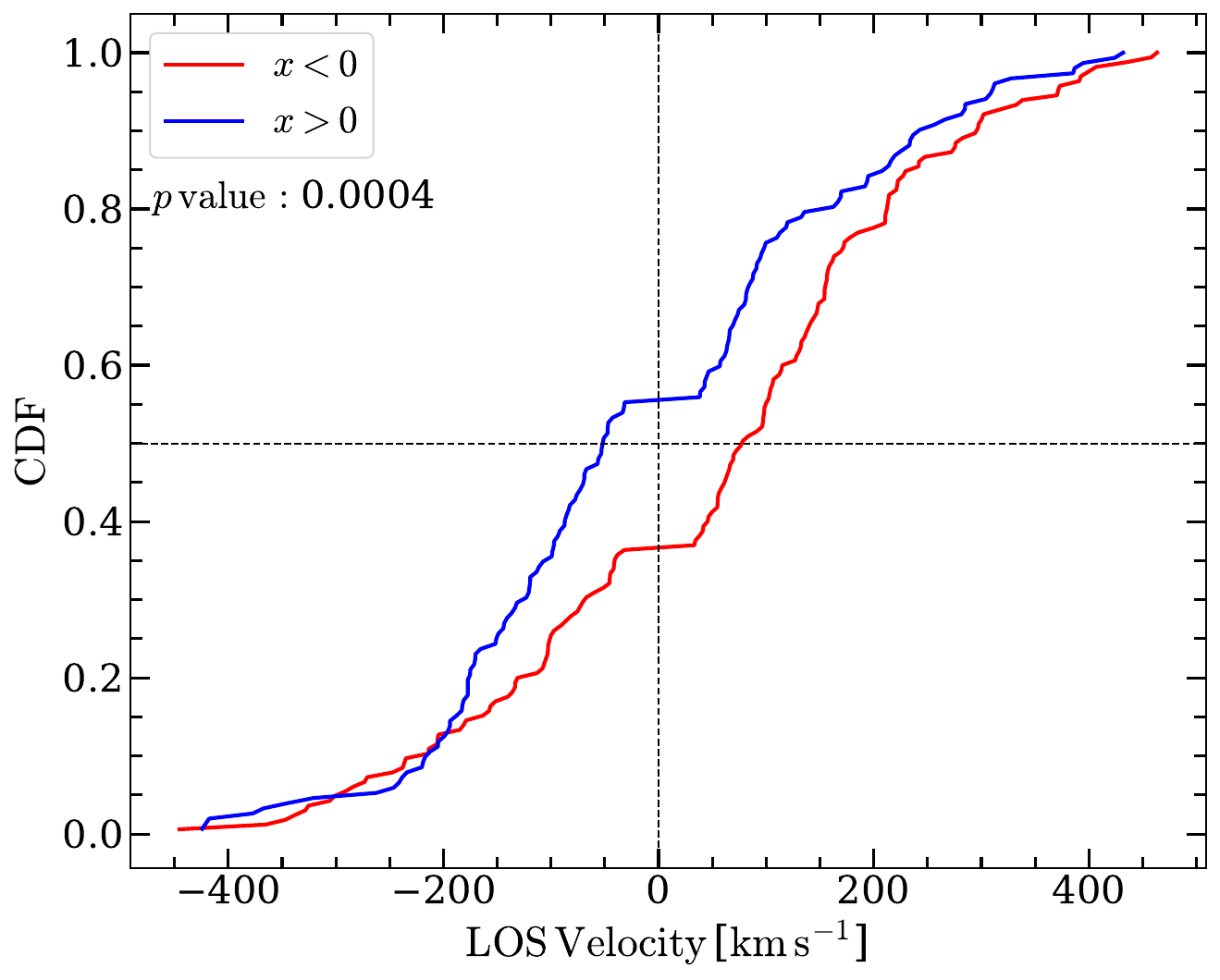}
\caption{The CDFs of the distributions in Fig.\,\ref{fig:LOS_distribution}. Red and blue lines are for the $x<0$ and $x>0$ species. Dashed vertical and horizontal lines represent the LOS velocity of zero and CDF value of 0.5. Upper left is the $p$-value of the K-S test between this two distributions. The platforms of the two distributions near the zero point are due to the cut of LOS velocities larger than $30$ km/s.}
\label{fig:K-S_test_LOS}
\end{figure}

In addition to splitting the population to the the red-shifted and blue-shifted parts by their LOS velocities and examining the difference in their offset preferences, we also divided the satellites 
into two groups with $x>0$ and $x<0$ and compared their LOS velocity distributions. The result is shown in Fig.\,\ref{fig:LOS_distribution}. According to our stacking procedures in Sec.\,\ref{Stack}, $x>0$ and $x<0$ actually correspond to the blue-shifted and red-shifted parts of the stellar discs.  The K-S test shown in Fig.\,\ref{fig:K-S_test_LOS} indicates the two distributions are different at a confidence level of 99.96 percent ($\sim 3.5\sigma$, with $p$-value of 0.0004). The fact that the satellites with $x>0$ ($x<0$) exhibit a LOS velocity distribution toward the blue-shifted (red-shifted) case, same as that of the stellar disk rotation direction, again implies the co-rotation between the stellar discs and satellites.

\section{Conclusions and Discussions}
\label{Conclusions_Discussion}
In this paper, using two sets of observation data, i.e., the SDSS and MaNGA galaxy survey, and the GAMA and SAMI galaxy survey, we carry out a search for a coherent kinematic signal between the stellar discs and their ambient galaxies, as predicted in the TNG-100 simulation and explicitly demonstrated by  \citet{Lu_et_al_2022}.

To do so, we selected the late-type and S0 central galaxies that show clear patterns of regular stellar-disk rotation through visual inspection in the MaNGA and SAMI galaxy surveys, rotated and stacked the kinematic field of these systems (see Sec.\,\ref{Stack}) to search for the predicted coherent kinematics. A K-S test showed that the distributions of the projected locations of the red-shifted and the blue-shifted satellite populations differ at a confidence level of 99.7 percent ($\sim 3\sigma$, see Sec.\,\ref{Co-rotation} and Figs. \ref{fig:MaNGA_SAMI} and \ref{fig:K-S_test} therein). In particular, the difference is as such that the red-shifted (blue-shifted) satellites are preferentially distributed on the red-shifted (blue-shifted) side of the stellar discs of the host galaxies. We also ran random tests to our samples (see Fig.\,\ref{fig:K-S_test}), which shows the co-rotation can hardly be derived from random distribution. 
While the two populations that are divided according to their projected locations with respect to the central stellar disc also exhibit different LOS velocity distributions. The confidence level for such a difference is 99.96 percent ($\sim 3.5\sigma$, see Figs. \ref{fig:LOS_distribution} and \ref{fig:K-S_test_LOS}). The difference is present as such that satellite galaxies projected on the red-shifted (blue-shifted) of the rotating stellar discs also preferentially have red-shifted (blue-shifted) LOS velocities. All tests above confirm the prediction that the satellites are co-rotating with the central stellar discs.


The co-rotation signal, as is presented in Fig.\,\ref{fig:MaNGA_SAMI}, is significant at about 100 kpc ($\sim \pm 20\,R_{\rm eff}$) and becomes weaker when going to larger scales. According to the angular momentum acquisition and modulation scenario \cite{Wang_et_al_2022,Lu_et_al_2022}, the spin of the stellar disc is an inheritance of the cold CGM, the spin of which is further modulated by its ambient environment (satellites) through interactions like merging and fly-by. It is worth noting that the coherent kinematic signal is exactly the consequence of such an angular momentum inheritance from large to small scales. However, this can only be observed when stacking a sizeable sample of galaxies in their merging and interacting environment. On individual cases, the cold CGM gas is always in form of localized in-spiral streams (see Figure 4 and 8 of \citealt{Wang_et_al_2022}) rather than a butterfly-like accretion disc, which can only be seen after stacking (see Figure 9 of \citealt{Lu_et_al_2022}). Equally, the multi-direction galaxy accretion would not guarantee that we observe coherent rotations of satellites around each and every individual host. The coherent kinetic signal as reported in this work can only be seen after stacking systems lined up according to kinematic major axes of the stellar discs.   


Our discovery is generally consistent with previous observational studies on the co-rotation between galaxies (i.e., their stellar discs) and their environment. In \citet{Lee_et_al_2019a}, the author found clear evidence (3.5 $\sigma$) of dynamical coherence (measured by luminosity-weighted velocity difference in neighbouring galaxies that are projected on different kinematic sides of central galaxies) existing between galaxy \textit{outskirts} and their neighbours (\textit{not only satellites}), up to 800 kpc using the Calar Alto Legacy Integral Field Area (CALIFA, \citealt{Sanchez_et_al_2012}) survey data and NASA-Sloan Atlas (NSA, created by Michael Blanton) catalog. The coherent signal gets stronger at smaller distances within 300 kpc (see Figure 11 therein). 
\citet{Lee_et_al_2020} further showed that the small-scale dynamical coherence is also accompanied by galaxy conformity, indicating that galaxy interactions are responsible for both observations. In \citet{Lee_et_al_2019b},  the authors claimed that such a dynamical coherence can extend up to $\sim$ 6 Mpc (at 2.8 $\sigma$), which is still not well-explained. 
It is interesting to note that \citet{Mai_et_al_2022}, following a similar method as \citet{Lee_et_al_2019b}, 
however, did not detect similar coherent signals as  up to a projected distance of 10 Mpc from central galaxies. 
They also suggested that the coherence at larger scales might be caused by the coincidental scatter or the variance of large-scale structure rather than physical relations between galaxy spins and their neighbours. 

In general, galaxy systems are born in their large-scale environment. Their  evolution is also largely shaped by the large-scale field. The large-scale matter distributions of cosmic structures set up initial conditions for galaxies to be born with and also determine how materials like gas and gravitationally-bound haloes at smaller scales are accreted from nearby cosmic webs to galaxy centres. The former flows into a galaxy's center and forms stars, while the latter become satellites (or passers-by) and provide modulations to the galaxy through interactions. 
Here in this paper, we search for a co-rotation pattern among galaxy groups, which is only significant at scales of 100 kpc. It would be interesting to investigate how far this signal may extend to, as a consequence of co-evolution of galaxies in their large-scale environment.  

\section*{Acknowledgements}

We would like to thank Prof. Xiaohu Yang for his patience when we had troubles with the catalog data. We are also grateful to Yanhan Guo for his kind help on the explanation and data of volume corrections. Also, we want to thank Prof. Simon White for his constructive suggestions about the statistics on the kinematical coherence.

\section*{Data availability}
YangDR7 catalogues are available from \url{https://gax.sjtu.edu.cn/data/Group.html}. The other group and galaxy data are obtained from official websites of MaNGA, GAMA, and SAMI survey. The rest of the data used in the article will be shared when requests are applied reasonably.
 


\bibliographystyle{mnras}
\bibliography{reference} 




\appendix


\bsp	
\label{lastpage}
\end{document}